\documentclass[letter,twocolumn]{jpsj3}
\usepackage{graphicx,color}%
\begin{document}

\title{%
Electronic Structure of Novel Superconductor 
Ca$_4$Al$_2$O$_6$Fe$_2$As$_2$
}

\author{Takashi Miyake,$^{1,4,5}$ Taichi Kosugi,$^{1,3}$ Shoji Ishibashi$^{1,4}$ and Kiyoyuki Terakura$^{2}$}

\inst{$^1$Nanosystem Research Institute (NRI) "RICS", AIST, 
Umezono, Tsukuba 305-8568, Japan\\ 
$^2$Research Center for Integrated Science (RCIS),
JAIST
Asahidai, Nomi, Ishikawa 923-1292, Japan\\
$^3$Department of Physics, University of Tokyo, Bunkyo, Tokyo 113-0033, Japan\\
$^4$JST, Transformative Research-Project on Iron Pnictides (TRIP), Sanbancho,
Chiyoda, Tokyo 102-0075, Japan \\
$^5$ JST, Core Research for Evolutional Science and Technology (CREST), 
Honcho, Kawaguchi, Saitama 332-0012, Japan 
}

\recdate{\today}

\abst{
We have performed the first-principles electronic structure calculation for the 
novel superconductor Ca$_4$Al$_2$O$_6$Fe$_2$As$_2$ which has the smallest $a$ lattice
parameter and the largest As height from the Fe plane among the Fe-As superconductors. 
We find that one of the hole-like Fermi surfaces is missing around the $\Gamma$ point 
compared to the case of LaFeAsO. 
Analysis using the maximally-localized-Wannier-function technique indicates that
the $xy$ orbital becomes more localized as the As-Fe-As angle decreases. 
This induces rearrangement of bands, which results in 
the change of the Fermi-surface topology of 
Ca$_4$Al$_2$O$_6$Fe$_2$As$_2$ from that of LaFeAsO. 
The strength of electron correlation is also evaluated using the constraint RPA method, 
and it turns out that Ca$_4$Al$_2$O$_6$Fe$_2$As$_2$ is more correlated than LaFeAsO. 
}

\maketitle

\bigskip

The discovery of the new iron-pnictide superconductor LaO$_{1-x}$F$_x$FeAs\cite{kamihara}  stimulated
extensive work to clarify the basic features of material properties, particularly the origin and character of
superconductivity and also to search for new superconductors.  In fact, a variety of related superconductors
have been synthesized: they have the topologically identical iron (Fe)-pnictogen (Pn) layer structure combined 
with various block layers.\cite{rotter,tapp,shimizu}  Among them, a family of materials which have block layers of the perovskite
structure are unique in the sense that the distance between the neighboring Fe-Pn layers can be systematically
controlled by changing the thickness of the block layer.\cite{shimizu}  The rich variation of the materials with the perovskite
structure is also attractive.  Quite recently, Ogino {\em et al.} synthesized (Ca$_{n+2}$(Al,Ti)$_n$O$_{3n-y}$)(Fe$_2$As$_2$) ($n$=2,3,4)
and observed the start of resistivity drop at about 39 K for $n$=4.\cite{ogino} For the synthesis of these materials at ambient
pressure, the presence of Ti turns out to be essential not to produce strong internal strain.  However, subsequent to
this work, Shirage {\em et al.} have successfully synthesized   Ca$_4$Al$_2$O$_{6-y}$Fe$_2$Pn$_2$ with Pn=P and As
without Ti using a high-pressure synthesis technique.\cite{shirage}  The superconducting transition temperatures are
17.1 K for P and 28.3 K for As.  These materials are characterized by the smallest $a$-lattice parameters
and the largest pnictogen heights from the Fe plane among the iron-pnictide superconductors.
It is pointed out that these quantities are crucial in determining the transition temperature.\cite{lee,zhao,vildosola,kuroki}

In the present work, we have performed DFT calculations for (Ca$_4$Al$_2$O$_{6-y}$)(Fe$_2$As$_2$) in order to analyze
the effects of the above unique structural features on the electronic structures paying particular attention to the
energy range near the Fermi level.  The absence of Ti in this material makes our analysis simpler because d states near
the Fermi level comes only from Fe. 
We used a computational code QMAS \cite{qmas} 
based on the projector augmented-wave method\cite{blochl}, 
which has been applied to the study of the ground state properties 
of LaFeAsO\cite{ishibashi,ishibashi2} and SrFe$_2$As$_2$.\cite{ishibashi3} 
The generalized gradient approximation (GGA)\cite{perdew} was adopted 
for the electronic exchange correlation energy.  
The electronic band structure in the vicinity 
of the Fermi level is analyzed in detail 
using the maximally-localized Wannier function technique \cite{marzari,souza} 
combined with the full-potential linear muffin-tin orbital (FP-LMTO) method \cite{methfessel}
in the local density approximation (LDA). \cite{vwn} 

\begin{figure}[t]
\begin{center}
\includegraphics[width=7.cm,trim= 2.cm 1.5cm 2.cm 2.5cm]{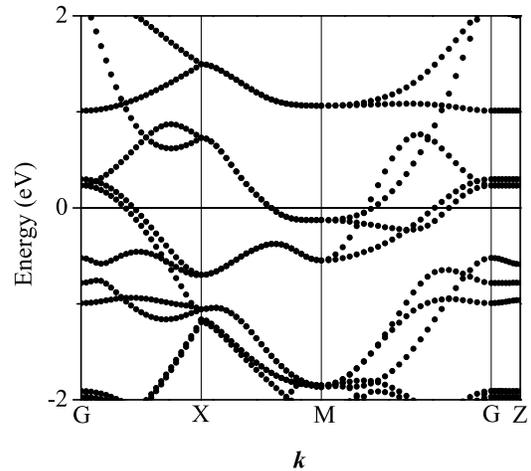}
\end{center}
\caption{Non-magnetic electronic band structure for 
Ca$_4$Al$_2$O$_6$Fe$_2$As$_2$. 
The Fermi level corresponds to the energy zero.
}
\label{f1}
\end{figure}

\begin{figure}[ht]
\begin{center}
\includegraphics[width=7.cm,trim= 2.cm .5cm 1.cm 0.cm]{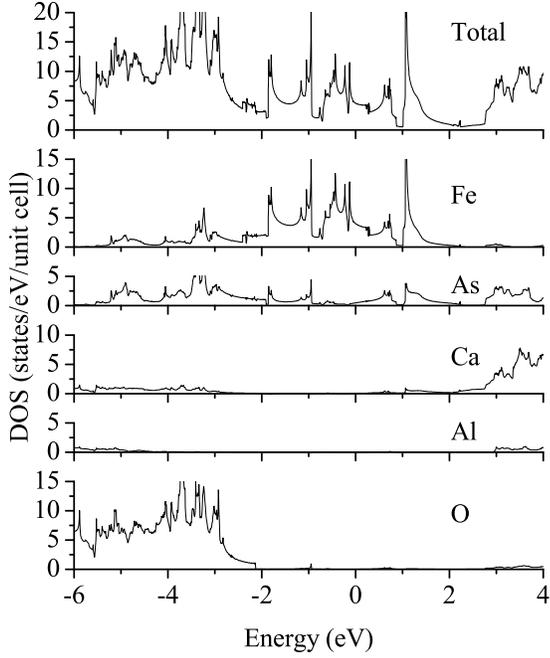}
\end{center}
\caption{Electronic density of states (DOS) and partial DOS's for 
Ca$_4$Al$_2$O$_6$Fe$_2$As$_2$. 
The Fermi level corresponds to the energy zero.
}
\label{f2}
\end{figure}

Figures 1 and 2 represent the non-magnetic electronic band structure and density of states 
for the experimental structure of 
Ca$_4$Al$_2$O$_6$Fe$_2$As$_2$ \cite{shirage} 
in GGA obtained by QMAS. 
We found that LDA (using the FP-LMTO code) gives almost the same result.
As with other iron-pnictide superconductors, electronic states 
near the Fermi level derive predominantly from the Fe 3d orbitals 
though there is significant contribution of As 4p states. 
In Fig. 3 (a), the Fermi surfaces are shown for the  undoped case and the doped cases
with $\pm$ 0.1 electrons/Fe obtained by the rigid band approximation. 
There are two electron-like surfaces centered at the M point. 
This is a common feature among the iron-pnictide superconductors. 
On the other hand, there are two hole-like surfaces around the $\Gamma$ point.  
This is in sharp contrast with other iron-pnictide superconductors, 
where three hole-like surfaces are formed. We will discuss this point later.  
To examine the nesting property, a bare generalized susceptibility is calculated and plotted in Fig. 3 (b).
There is a nesting feature from $\Gamma$ to M. 
While the change is small with hole doping, 
it is reduced significantly with electron doping, suggesting that 
the magnetic instability is suppressed under electron doping.
In the actual samples, oxygen content $6-y$=5.6 to 5.8 will produce doping of 0.2 to 0.4 electrons/Fe. 

\begin{figure}[ht]
\begin{center}
\includegraphics[width=8.cm,trim= 0.cm 0.cm 0.cm 0.cm]{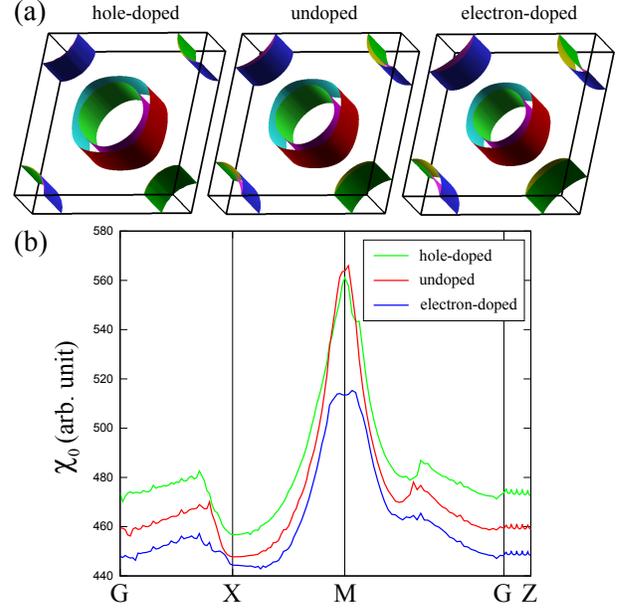}
\end{center}
\caption{(Color) 
(a) Fermi surfaces for hole-doped (left), undoped (middle) and electron-doped (right) Ca$_4$Al$_2$O$_6$Fe$_2$As$_2$, 
(b) Bare generalized susceptibility for these Fermi surfaces.}
\label{f3}
\end{figure}

\begin{figure}[b]
\begin{center}
\includegraphics[width=9.cm,trim= 0.cm .0cm 0.cm 0.3cm]{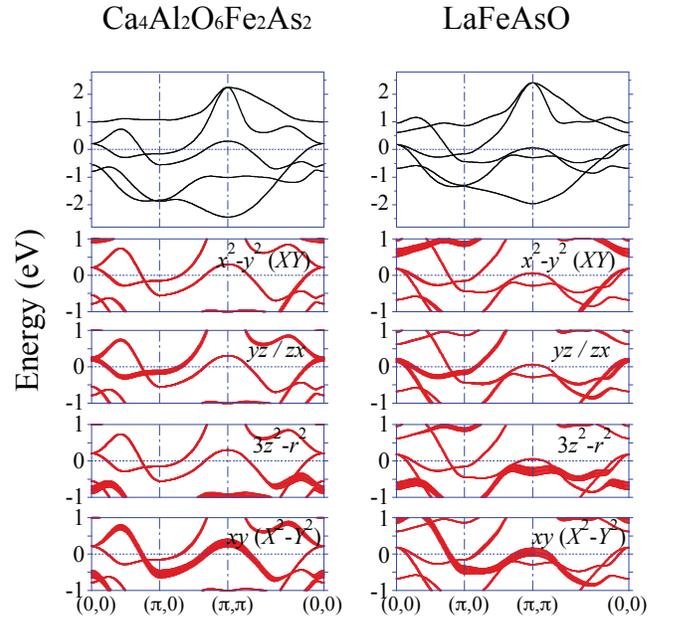}
\end{center}
\caption{(Color online)
Unfolded band structure of Ca$_4$Al$_2$O$_6$Fe$_2$As$_2$ (left) and LaFeAsO (right). 
The width of the lines represent the weight of each maximally localized Wannier function. 
The Fermi level corresponds to the energy zero. 
The $x$ and $y$ refer to the unit cell containing one iron atom in the cell, 
whereas $XY$ axis is rotated by 45 degrees, corresponding to the original cell.}
\label{f4}
\end{figure}

In order to analyze the electronic structure in more detail, 
we projected out energy bands near the Fermi energy 
and constructed ten maximally localized Wannier functions. 
The Wannier functions are characterized by the shapes closely related 
to the Fe 3d orbitals though there are significant amplitudes at adjacent As atoms. 
This is consistent with the feature of PDOS's mentioned above. 
The interpolated band structure is then unfolded following the procedure of Ref.~\ref{unfold}.
The result is shown and compared with that for LaFeAsO in Figure \ref{f4}. 
Here the width of the lines represents the weight of each Wannier function component. 
In the $yz/zx$ panel, the weight is defined as the average of the weight of $d_{yz}$ and that of $d_{zx}$. 
Our convention for the $x$ and $y$ axes is the same as Fig.1 in Ref.~\ref{unfold}. 
Namely, we define the $x$ and $y$ for the unit cell containing one Fe atom, 
whereas the $X$ and $Y$ refer to the original cell, which is rotated by 45 degrees 
from the $x$ and $y$ axes.

\begin{figure}[b]
\begin{center}
\includegraphics[height=5cm,width=9.cm,trim= 0.cm .0cm 0.cm 1.cm]{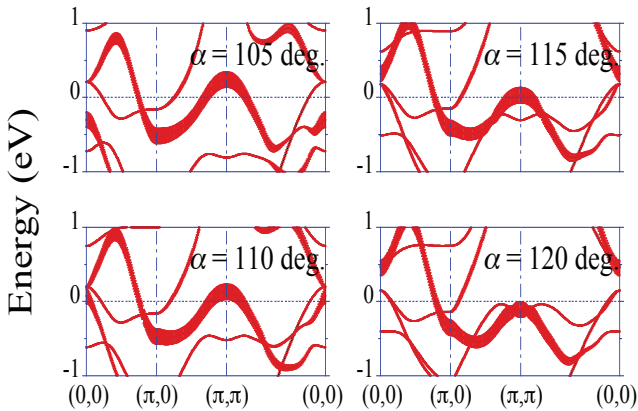}
\end{center}
\caption{(Color online)
Unfolded band structure of Ca$_4$Al$_2$O$_6$Fe$_2$As$_2$ 
for the As-Fe-As angle $\alpha$ =105, 110, 115 and 120 degrees. 
The width of the lines represents the weight of the $xy$ Wannier function. 
See the caption of Fig.\ref{f4} for the convention of the $xy$ axis. 
}
\label{f5}
\end{figure}

There are obvious differences in the band structure near (0,0) and ($\pi,\pi$) points
between the two compounds in Fig.\ref{f4}.
In order to understand the origins of these differences, 
we calculated the band structure of Ca$_4$Al$_2$O$_6$Fe$_2$As$_2$ 
for different values of $\alpha$ while fixing both the Fe-As bond length and 
the distance between the As plane and the block layer. 
The band structure is shown in Fig.\ref{f5}, where the width of the lines represents 
the weight of the $xy$ character. 
We can see that $\alpha$ does affect the band structure significantly. 
At $\alpha$=105 deg., a state appears at $-$0.3 eV at (0,0). This is basically the case
of Ca$_4$Al$_2$O$_6$Fe$_2$As$_2$ whose $\alpha$ is 102.13 degrees.
As $\alpha$ increases, the state moves up and forms the hole-like Fermi surface in the 
case of $\alpha$=110 deg. With further increase of $\alpha$, the band structure becomes similar to that of
LaFeAsO whose $\alpha$ is 113.55 degrees. 
The behavior of the $xy$ band near ($\pi,\pi$) with the change of $\alpha$ or the pnictogen height
from the Fe plane is qualitatively the same as what was explained by Kuroki et al.\cite{kuroki}.
The $xy$ band moves up above 
the Fermi level as $\alpha$ decreases and forms a hole pocket near ($\pi,\pi$), whose
presence may be a crucial condition for the strong superconductivity of s$\pm$ character.
We also found that the distance  between FeAs layer and block layer is another important parameter. 
With increasing the distance, the band rearrangement takes place at a larger value of $\alpha$. 

\begin{figure}[ht]
\begin{center}
\includegraphics[width=9.cm,trim= 0.cm .0cm 0.cm 0.5cm]{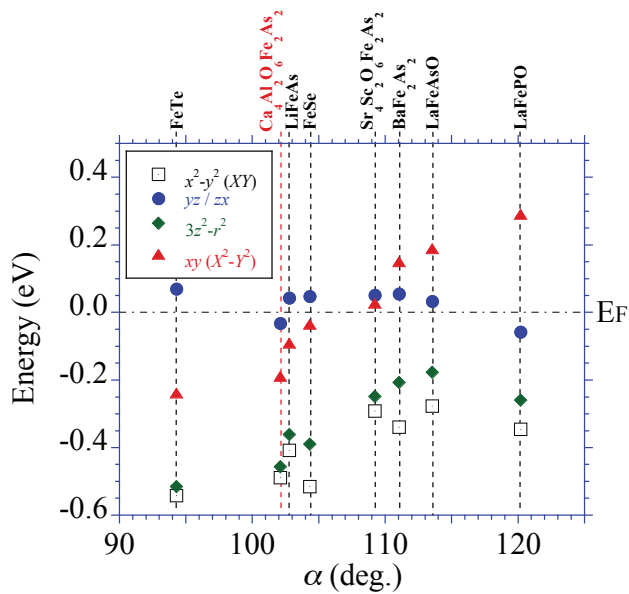}
\end{center}
\caption{(Color online)
Onsite energy of each Wannier function for eight compounds 
as a function of the $X$-Fe-$X$ angle, where $X$ is the pnictogen / chalcogen atom. 
The Fermi level corresponds to the energy zero. 
See the caption of Fig.\ref{f4} for the convention of the $xy$ axis.}
\label{f6}
\end{figure}

As suggested by the above arguments, the $xy$ ($X^2-Y^2$) orbital plays an important role 
in the band rearrangement. For the analysis of the behavior of this band, we list in Table~\ref{t1} 
the onsite energy and the
hopping integrals between $xy$ ($X^2-Y^2$) orbitals up to 5th neighbors, beyond which
the hopping integrals are negligibly small. 
We note that the sign of $t_1$ and $t_4$ depends on the choice of the gauge. 
We take the opposite sign compared to Ref.~\ref{miyake_jpsj10}. 
By cutting off the hopping integrals between different kinds of d orbitals,
the band dispersion for the $xy$ ($X^2-Y^2$) orbitals is written as
\begin{eqnarray}
E_{xy}({\bf k}) & = & \epsilon_{xy} + 2t_1 [ \cos(k_x) + \cos(k_y) ]  \nonumber \\
& & + 4 t_2 \cos(k_x)\cos(k_y) \nonumber \\
& & +2t_3 [\cos(2k_x) + \cos(2k_y)] \nonumber  \\
& & +4t_4 [\cos(2k_x)\cos(k_y) + \cos(k_x)\cos(2k_y) ] \nonumber \\
& & +4t_5 \cos(2k_x)\cos(2k_y) \;,
\label{e1}
\end{eqnarray}
where $\epsilon_{xy}$ is the onsite energy, and $t_n$ is the hopping integral for the $n$th neighbors. 
Now we can clearly understand why the $xy$ band behaves quite differently between 
LaFeAsO and Ca$_4$Al$_2$O$_6$Fe$_2$As$_2$.  First of all, the onsite energy of Ca$_4$Al$_2$O$_6$Fe$_2$As$_2$ is
about 0.4 eV deeper than that of LaFeAsO (Table~\ref{t1} and Fig.\ref{f6}) because of the change in the crystal field as implied by the difference in $\alpha$.
Secondly, the hopping integrals for the first three neighbors are quite different between the two compounds.
Particularly the first neighbor hopping of Ca$_4$Al$_2$O$_6$Fe$_2$As$_2$ is very small in magnitude and has different sign
from that of LaFeAsO.  In Table~\ref{t1}, the numbers in the parenthesis for $t_1$ denote the contributions
from the direct d-d hopping.  In the LaFeAsO case, the contribution from the indirect Fe-As-Fe
hopping (0.394 eV) overwhelms the direct hopping ($-$0.243 eV) and the net hopping integral is 0.151 eV.
However, in the Ca$_4$Al$_2$O$_6$Fe$_2$As$_2$ case, the indirect contribution (0.329 eV)
and the direct contribution ($-$0.369 eV) nearly cancel each other to produce a small net
hopping integral of $-$0.04 eV.  The larger direct hopping for Ca$_4$Al$_2$O$_6$Fe$_2$As$_2$ is
due to smaller Fe-Fe distance and the smaller indirect hopping is due to smaller $\alpha$.  

Using Eq. \ref{e1} and the parameters in Table~\ref{t1}, we found that not only the difference in the onsite energy 
but also hopping integrals produce a large difference in the band energy at (0,0) between
the two compounds.  In addition to this, 
the difference in the
band energies between (0,0) and ($\pi,\pi$) 
is $E_{xy}(\pi,\pi)-E_{xy}(0,0)= - 8(t_1 + 2t_4)$.
Actual values for this quantity are $-$0.712 eV for LaFeAsO and +0.848 eV 
for Ca$_4$Al$_2$O$_6$Fe$_2$As$_2$. 
One may also notice in Fig.\ref{f4} that the position of the $3z^2-r^2$ branch near ($\pi,\pi$) is quite different
between the compounds.  For this aspect, we only mention that the trend in the onsite energy shown 
in Fig.\ref{f6} is also enhanced by the hopping integrals as discussed for the $xy$ $(X^2-Y^2)$ branch.

Now we discuss briefly other aspects of the band structure.
As for the $x^2-y^2$ ($XY$) and $yz/zx$ states, the hopping integrals are 
qualitatively the same between LaFeAsO and Ca$_4$Al$_2$O$_6$Fe$_2$As$_2$.  
However, the variation in the onsite energies shown in Fig.\ref{f6}
may produce qualitatively important effects on the electronic structure of these states. 
In Fig.\ref{f6}, we first observe
the splitting between the d$\epsilon$($xy$, $yz$, $zx$) states and the d$\gamma$($x^2-y^2$, 3$z^2-r^2$) states 
when the tetragonal distortion is small ($\alpha$$\sim$109.4 deg.) 
in accordance with the ligand field theory for the tetrahedral symmetry.
Further splitting among states within each of d$\epsilon$ and d$\gamma$
as $\alpha$ deviates from 109.4 deg. is a natural consequence of the distortion in
the tetrahedral coordination.
We also observe a quite interesting fact that
the onsite energy of $yz/zx$ states with respect to the Fermi level is nearly invariant 
for a wide range of materials with different $\alpha$s.
Other states particularly of d$\gamma$ become deeper and the energy
separation from $yz/zx$ states becomes larger as $\alpha$ decreases.  This variation
in the onsite energies leads to the increase of
the weight of d$\epsilon$ states near the Fermi level as already pointed out by
our previous papers.\cite{ishibashi3,miyake10}

\begin{table}
\caption{
Orbital energy and hopping integrals up to fifth neighbors for $xy$ ($X^2-Y^2$) state (in eV).  
The notation with the $X$ and $Y$ axes is for easier comparison with 
the previous works on different compounds \cite{nakamura,miyake10}. 
See the caption of Fig.\ref{f4} for the convention of $X$ and $Y$ axes. 
For $t_1$, 
the numbers in parenthesis denote the contribution from the direct d-d hopping estimated by the dpp model.
}
\begin{tabular}{lccc} 
\hline \hline \\ [-8pt]  
 & $\varepsilon_{xy}$ & $t_1$ & $t_2$ \\ 
\hline \\ [-8pt] 
LaFeAsO & 0.189 & 0.151 ($-$0.243) & 0.117 \\ 
Ca$_4$Al$_2$O$_6$Fe$_2$As$_2$ & $-$0.189 & $-$0.040 ($-$0.369) & 0.066 \\
\hline \\ [-8pt]  
 & $t_3$ & $t_4$ & $t_5$ \\ 
\hline \\ [-8pt] 
LaFeAsO & $-$0.024 & $-$0.031 & $-$0.027 \\
Ca$_4$Al$_2$O$_6$Fe$_2$As$_2$ & $-$0.012 & $-$0.033 & $-$0.029 \\
\hline 
\hline 
\end{tabular}
\label{t1}
\end{table} 

Finally we mention the strength of electron correlation. 
The onsite effective Coulomb interaction in the constraint random phase approximation 
(cRPA) \cite{aryasetiawan04,miyake_crpa} are shown in Table~\ref{t2}. 
The results of other iron-based superconductors using the same technique can be 
found in Ref.~\ref{miyake_jpsj10},\ref{nakamura_jpsj08},\ref{miyake_jpsj08}.
The average of the diagonal terms 
is 3.08 eV, which 
is larger than that of LaFeAsO (2.53 eV) and substantially smaller 
than that of FeSe (4.24 eV). \cite{miyake10} 
The exchange energy (not shown) is smaller than that of FeSe, and 
very close to that of LaFeAsO, with a slightly larger value for the $xy$ orbital. 
The orbital anisotropy is weaker than LaFeAsO, 
reflecting that the $xy$ ($X^2$-$Y^2$) orbital is less extended. 

\begin{table} 
\caption{
Effective Coulomb interaction ($U$) 
on the same iron site for all the combinations of Wannier functions in units of eV. 
The definitions of $X$ and $Y$ axes are the same as those in Table~\ref{t1}.
}
\begin{tabular}{cccccc} 
\hline \hline \\ [-8pt]  
 & $XY$& $YZ$ & $3Z^2-r^2$ & $ZX$ &  $X^2-Y^2$ \\
\hline \\ [-8pt] 
$XY$ & 3.38 & 2.21 & 2.24 & 2.21 & 2.53  \\
$YZ$ & 2.21 & 2.87 & 2.45 & 2.03 & 2.07  \\
$3Z^2-r^2$ & 2.24 & 2.45 & 3.46 & 2.45 & 2.11 \\
$ZX$ & 2.21 & 2.03 & 2.45 & 2.87 & 2.07  \\
$X^2-Y^2$ & 2.53 & 2.07 & 2.11 & 2.07 & 2.82  \\ 
\hline 
\hline 
\end{tabular} 
\label{t2}
\end{table} 

In conclusion, we have studied the electronic structure of 
Ca$_4$Al$_2$O$_6$Fe$_2$As$_2$. 
We found that the topology of the Fermi surface is different 
from that of LaFeAsO. 
It is revealed that the As-Fe-As angle has strong impact on 
the band structure in the vicinity of the Fermi level around the (0,0) point, 
specifically through the change in the onsite energy level and transfer integral 
of the $xy$ orbital. 
It would be interesting to see how the pressure changes the properties of the compound. 
Relation with superconductivity is also an open question.
Magnetic calculations including structural optimization are 
ongoing and will be published elsewhere.

The authors are grateful to Dr. P. M. Shirage, Dr. K. Kihou, Dr. C.-H. Lee, Dr. H. Kito, 
Dr. H. Eisaki and Dr. A. Iyo for providing us with experimental data prior to publication. 
The authors  also acknowledge Prof. K. Kuroki for discussions. 
The present work is partially supported by the Next Generation Supercomputer Project, 
Nanoscience Program, and by a Grant-in-Aid for Scientific Research on Innovative Areas 
under grant No. 22104010  from MEXT, Japan.

\end{document}